\documentclass[12pt]{article}
\usepackage{graphicx}

\newcommand\pubnumber{CIPANP2018-Dunlop}
\newcommand\pubdate{\today}

\def\guelph{Department of Physics\\
 University of Guelph, Guelph, Ontario N1G 2W1, Canada}
 
\def\sfu{Department of Chemistry\\ Simon Fraser University, Burnaby, British Columbia V5A 1S6, Canada}

\def\triumf{TRIUMF, 4004 Wesbrook Mall, Vancouver, British Columbia V6T 2A3, Canada
}

\def\Title#1{\begin{center} {\Large #1 } \end{center}}
\def\Author#1{\begin{center}{ \sc #1} \end{center}}
\def\Address#1{\begin{center}{ \it #1} \end{center}}

\newcommand\pubblock{\rightline{\begin{tabular}{l} \pubnumber\\
         \pubdate  \end{tabular}}}
\newenvironment{Abstract}{\begin{quotation}  }{\end{quotation}}
\newenvironment{Presented}{\begin{quotation} \begin{center} 
             PRESENTED AT\end{center}\bigskip 
      \begin{center}\begin{large}}{\end{large}\end{center} \end{quotation}}

\def\beq{\begin{equation}}
\def\eeq#1{\label{#1}\end{equation}}
\def\eeqn{\end{equation}}

\def\beqa{\begin{eqnarray}}
\def\eeqa#1{\label{#1}\end{eqnarray}}
\def\eeqan{\end{eqnarray}}

\let\bar=\overbar

\def\Dslash{\not{\hbox{\kern-4pt $D$}}}
\def\dslash{\not{\hbox{\kern-2pt $\del$}}}

\def\msb{{\bar{\ssstyle M \kern -1pt S}}}

\begin{document}
\begin{titlepage}
\pubblock

\vfill
\Title{Half-Lives of Neutron Rich $^{130}$Cd and $^{131}$In}
\vfill
\Author{R. Dunlop, C.~E.~Svensson, V.~Bildstein, H.~Bidaman, P. Boubel, C.~Burbadge, M.~R.~Dunlop, P.~E.~Garrett, A.~D.~MacLean, E.~McGee, D.~M\"{u}cher, A.~J.~Radich, J.~Turko, T.~Zidar}
\Address{\guelph}

\Author{C.~Andreoiu, F.~H.~Garcia, K.~Ortner, J.~L.~Pore,  K.~Whitmore}
\Address{\sfu}

\Author{G.~C.~Ball, N.~Bernier, R.~Caballero-Folch, I.~Dillmann, L.~J.~Evitts, A.~B.~Garnsworthy, G.~Hackman, S.~Hallam, J.~Henderson, R.~Kr\"{u}cken, J. Lassen, R.~Li, E.~MacConnachie, M.~Moukaddam, B.~Olaizola, J.~Park, O.~Paetkau, P.~Ruotsalainen, J.~Smallcombe, J.~K.~Smith, A.~Teigelh{\"o}fer  }
\Address{\triumf}

\Author{A.~Jungclaus}
\Address{Instituto de Estructura de la Materia\\ CSIC, E-28006 Madrid, Spain}

\Author{S.~Ilyushkin}
\Address{Department of Physics\\ Colorado School of Mines, Golden, Colorado 80401, USA}

\Author{E.~Padilla-Rodal}
\Address{Universidad Nacional Aut\'onoma de M\'exico\\ Instituto de Ciencias Nucleares, AP 70-543, M\'exico City
04510, DF, M\'exico}

\vfill
\begin{Abstract}
The half-lives of isotopes around the $N=82$ shell closure are an important ingredient in astrophysical simulations and strongly influence the magnitude of the second $r$-process abundance peak in the $A\sim130$ region. The most neutron-rich $N=82$ nuclei are not accessible to the current generation of radioactive beam facilities and $r$-process simulations must therefore rely on calculations of the half-lives of the isotopes involved. Half-life measurements of the experimentally accessible nuclei in this region are important in order to benchmark these calculations. The half-life of $^{130}$Cd is particularly important as it is used to tune the Gamow-Teller quenching in shell-model calculations for the $\beta$ decay of other nuclei in this region. In this work, the GRIFFIN $\gamma$-ray spectrometer at the TRIUMF-ISAC facility was used to measure the half-life of $^{130}_{~48}$Cd$_{82}$ to be $T_{1/2}= 126(4)$~ms. In addition, the half-lives of the three $\beta$ decaying states of $^{131}_{~49}$In$_{82}$ were measured to be $T_{1/2}(1/2^-)=328(15)$~ms, $T_{1/2}(9/2^+)=265(8)$~ms, and $T_{1/2}(21/2^+)=323(50)$~ms, respectively, providing an important benchmark for half-life calculations in this region.


\end{Abstract}
\vfill
\begin{Presented}
Thirteenth Conference on the Intersections of Particle and Nuclear Physics\\
Palm Springs, California May 29--June 3, 2018
\end{Presented}
\vfill
\end{titlepage}
\def\thefootnote{\fnsymbol{footnote}}
\setcounter{footnote}{0}

\section{Introduction}
The astrophysical $r$-process is known to be responsible for the production of nearly half of the observed isotopes heavier than Fe~\cite{bbfh57,cameron57,cameron57-2}. In the $r$-process, the $N=82$ isotones act as waiting-points where an accumulation of $r$-process material occurs before it can $\beta$-decay to the next elemental chain. Furthermore, the relative half-lives of these nuclei determine the amount of material that is accumulated at each waiting point, and hence, the amplitude and shape of the resulting $r$-process abundance peaks~\cite{mumpower16,seeger65,cameron1983,cameron1983-2}. In the more neutron-rich ``cold'' $r$-process scenarios like neutron star mergers
\cite{freiburghaus99,korobkin12}, the reaction path is driven towards the neutron dripline. These regions are only partially accessible at the newest generation of radioactive beam facilities and calculations of the $r$-process flow must rely on the predictions of properties such as the half-lives, neutron separation energies, and neutron capture cross-sections. Therefore, to accurately model the $r$-process, it is critical to have models that can accurately reproduce these nuclear properties.   

In particular, shell-model calculations for the half-lives of nuclei at the $N=82$ neutron shell closure~\cite{cuenca07,zhi13} were performed by adjusting the quenching of the Gamow-Teller (GT) operator in order to reproduce the $^{130}_{~48}$Cd$_{82}$ half-life reported in Ref.~\cite{hannawald00}. These calculations produced half-lives that agree well with the experimental measurements for $^{130}$Cd~\cite{hannawald00} and $^{131}$In~\cite{PhysRevC.70.034312}, but are known to yield systematically large values for the half-lives of other $N=82$ nuclei~\cite{zhi13}. An experimental campaign at EURICA measuring 110 half-lives important to the $r$-process reported a significantly smaller half-life for $^{130}$Cd of~127(2)~ms~\cite{lorusso15}. A re-scaling of the GT quenching by a constant factor to reproduce this new $^{130}$Cd half-life resolves the discrepancy between the calculated and measured half-lives for the majority of the $N=82$ isotopes. However, this re-scaling creates a new discrepancy in the calculated half-life of $^{131}$In. In order to be confident in the half-life predictions of the neutron-rich $r$-process waiting-point nuclei, further experimetal and theoretical investigations are required.

\section{Experiment}
Gamma-Ray Infrastructure For Fundamental Investigations of Nuclei (GRIFFIN)~\cite{garnsworthy18,garnsworthy17,svensson14} is a high-efficiency $\gamma$-ray spectrometer comprised of 16 high-purity germanium (HPGe) clover detectors~\cite{RIZWAN2016126} located at the Isotope Separator and ACcelerator (ISAC) facility at TRIUMF~\cite{dilling13}. The $^{130}$Cd and $^{131}$In isotopes were each produced during separate beam times by impinging a 480-MeV proton beam with 9.8-$\mu$A intensity from the TRIUMF main cyclotron onto a UC$_x$ target. The isotopes of interest were selectively laser-ionized in a three-step excitation scheme while the ion-guide laser ion source (IG-LIS)~\cite{raeder14} was used to suppress surface-ionized isobars. A high-resolution mass separator ($\delta M/M \sim 1/2000$)~\cite{bricault13} was used to mass-select the isotope of interest. The low-energy (30~keV) radioactive ion beam (RIB) was then delivered to the GRIFFIN spectrometer in the \mbox{ISAC-I} experimental hall. Once at the GRIFFIN spectrometer, the RIB was implanted into a mylar tape at the mutual centers of SCEPTAR, an array of 20 thin plastic scintillators for tagging $\beta$ particles~\cite{ball05}, and GRIFFIN~\cite{garnsworthy18,garnsworthy17,svensson14}. 

The mylar tape allows for the experiment to be run in ``cycles-mode'' in order to optimize the experimental conditions. The longer-lived background activity, either from isobaric contaminants in the beam or from daughters following the decay of the isotope of interest, could be removed from the spectrometer by moving the tape, and therefore the beam spot, into a lead-shielded collection box following a measurement. For example, a typical cycle consisted of a background measurement for 0.5~s, followed by a collection period (beam-on) with the beam being implanted into the tape, followed by another data collection period (beam-off) with the beam blocked by the ISAC electrostatic beam kicker downstream of the mass separator. The tape was then moved over a period of 1~s to a shielded position outside of the array, completing the cycle and starting a new background measurement for the next cycle. The data used for determining the half-lives described below were recorded during the beam off, ``decay'' portion of the cycle.
\begin{figure}[!t]
    \centering
   \includegraphics[width=0.9\linewidth]{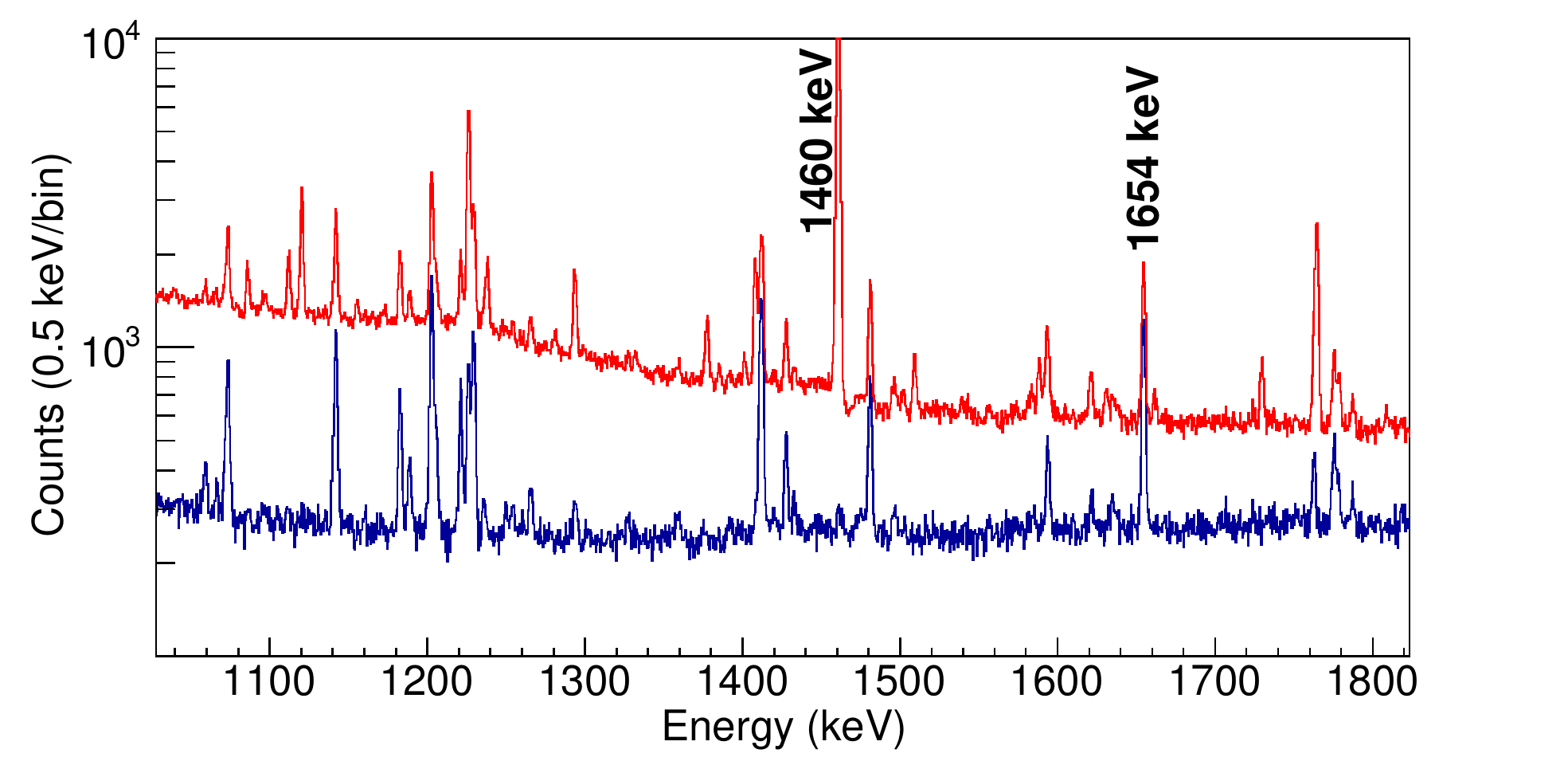}%
   \caption{(Color online) Comparison of (Red) $\gamma$ singles with (Blue) $\beta$-$\gamma$ coincidence spectra during the $^{131}$In experiment. In this region, the $\beta$-$\gamma$ coincidence leads to a strong suppression of room-background $\gamma$ rays as well as almost an order of magnitude reduction in the total Compton background. For example, the room-background 1460~keV ($^{40}$K) line is shown to be heavily suppressed, while the 1654~keV line in $^{131}$Sn has an increased peak-to-total ratio.
   \label{fig:beta-gamma}}
\end{figure}
The GRIFFIN HPGe data were analyzed using a clover addback algorithm in which all of the energy deposited into a HPGe clover within a coincidence window of 250~ns was summed. Using addback, Compton scattered $\gamma$-ray events that had full energy deposition within a single clover are recovered, increasing both the photopeak efficiency and the peak-to-total ratio in the $\gamma$-ray spectrum~\cite{garnsworthy18}.

The SCEPTAR array was used to suppress room-background $\gamma$-ray events by tagging $\beta$-decay events originating on the mylar tape, significantly increasing the peak-to-total ratio as shown in Fig.~\ref{fig:beta-gamma}. 

For the half-life measurements, sources of systematic uncertainties were investigated, including the re-binning of the data as well as a ``chop analysis''~\cite{grinyer05, laffoley13}. By changing the first bin included in the fit region, rate-dependent effects such as dead-time and pile-up would manifest as a correlation between the first bin used and the half-life. No statistically significant rate-dependent systematic effects were observed in either the $^{130}$Cd or $^{131}$In half-life measurements.

\section{Results}
\subsection{$^{130}$Cd Decay}\label{sec:cddecay}
During each $^{130}$Cd experimental cycle, a RIB containing 15-30~pps of $^{130}$Cd was implanted for 10~s, before the upstream electromagnetic kicker deflected the beam for 1.3~s. The tape was then moved, and the cycle was repeated. Approximately 38 hours of $^{130}$Cd data were collected. 

\begin{figure}[!t]
\centering
   \includegraphics[width=0.9\linewidth]{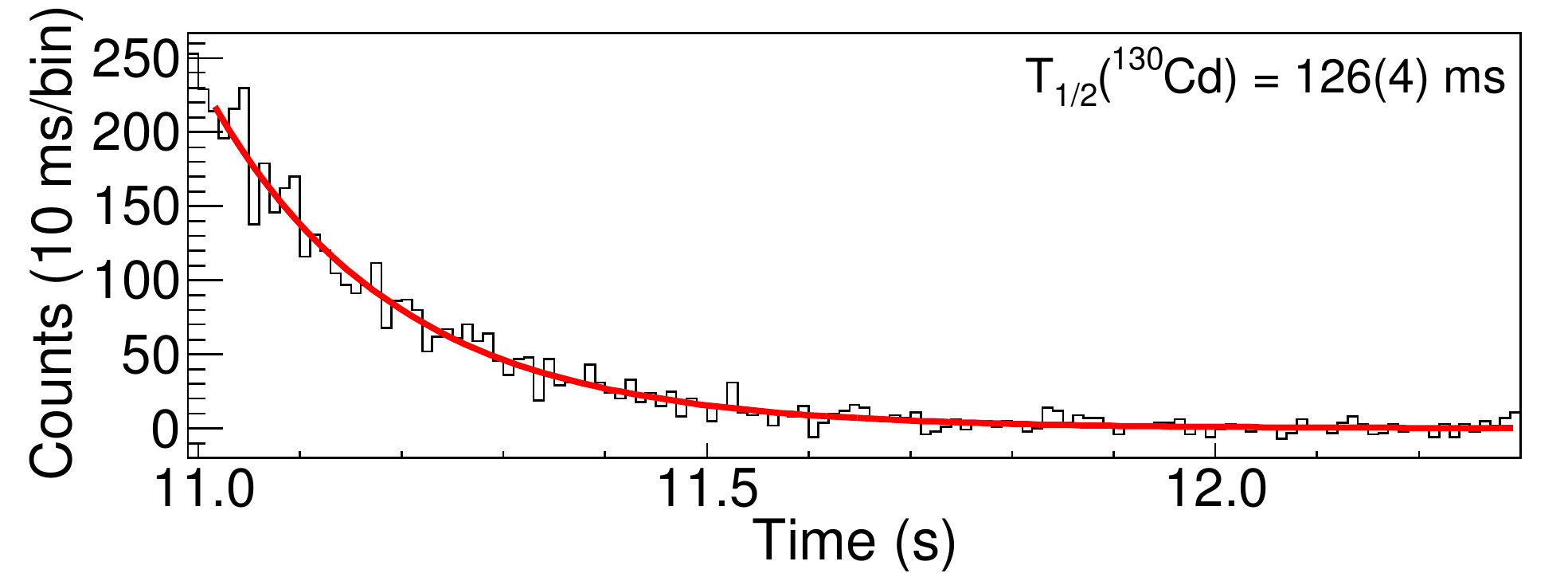}%
   \caption{Sum of the activity of the 451, 1170 and 1669~keV $\gamma$-rays in $^{130}$In following the $\beta$ decay of $^{130}$Cd during the decay portion of the cycle. The half-life obtained from the fit is 126(4)~ms~\cite{PhysRevC.93.062801}, in good agreement with the value of 127(2)~ms reported in Ref.~\cite{lorusso15} and in strong disagreement with the previous half-life measurement of 162(7)~ms~\cite{hannawald00}. \label{fig:Cd130hl}}
\end{figure}

The 451.0, 1170.3 and 1669.2 keV $\gamma$-rays following the decay of $^{130}$Cd~\cite{dillmann03} were used to measure the half-life. Figure~\ref{fig:Cd130hl} shows the fit to the activity of these $\gamma$ rays, yielding $T_{1/2}=126(4)$~ms for the decay of $^{130}$Cd~\cite{PhysRevC.93.062801}, in good agreement with the value of 127(2)~ms reported in Ref.~\cite{lorusso15} and in strong disagreement with the previous half-life measurement of 162(7)~ms~\cite{hannawald00}. 

The confirmation of the shorter half-life of $^{130}$Cd reported in Ref.~\cite{lorusso15} is important as it resolves many of the calculated half-life discrepancies for the neutron-rich $N=82$ nuclei. However, the reduced quenching of the GT operator implied by this result, creates a new discrepancy due to the newly reduced calculated half-life of $^{131}$In. Therefore, a spectroscopic study of the $\beta$ decay of $^{131}$In as well as a new measurement of the half-life was required in order to investigate the source of this disagreement between theory and experiment.

\subsection{$^{131}$In Decay}
The measurement of the half life of $^{131}$In is challenging due to the presence of three $\beta$-decaying states with similar half lives. A mixture of these three states, with spin assignments of $1/2^-$, $9/2^+$ and $21/2^+$~\cite{PhysRevC.70.034312}, was delivered to GRIFFIN. The high photo-peak efficiency of GRIFFIN made it possible to assign many of the $\gamma$-rays to a specific cascade via $\gamma$-$\gamma$ coincidences~\cite{dunlop18}. 

During each $^{131}$In experimental cycle, a RIB containing approximately 500 ions/s for each of the $9/2^+$ and $1/2^-$ isomers, and 60~ions/s for the $21/2^+$ isomer was implanted into the tape for 30~s, followed by a decay period of 5~s. The results reported here were obtained from 182 cycles recorded over a period of approximately 1.9~hours.
\subsubsection{Decay of the $21/2^+$ State}
The high-spin $21/2^+$ isomer of $^{131}$In decays to excited, high-spin states of $^{131}$Sn that deexcite to the $11/2^-$ isomer in $^{131}$Sn. No transitions connecting these high-spin states to the lower spin states of $^{131}$Sn that are populated following the $\beta$ decay of the $1/2^-$ and $9/2^+$ states of $^{131}$In have been reported thus far.

\begin{figure}[!t]
\centering
   \includegraphics[width=0.9\linewidth]{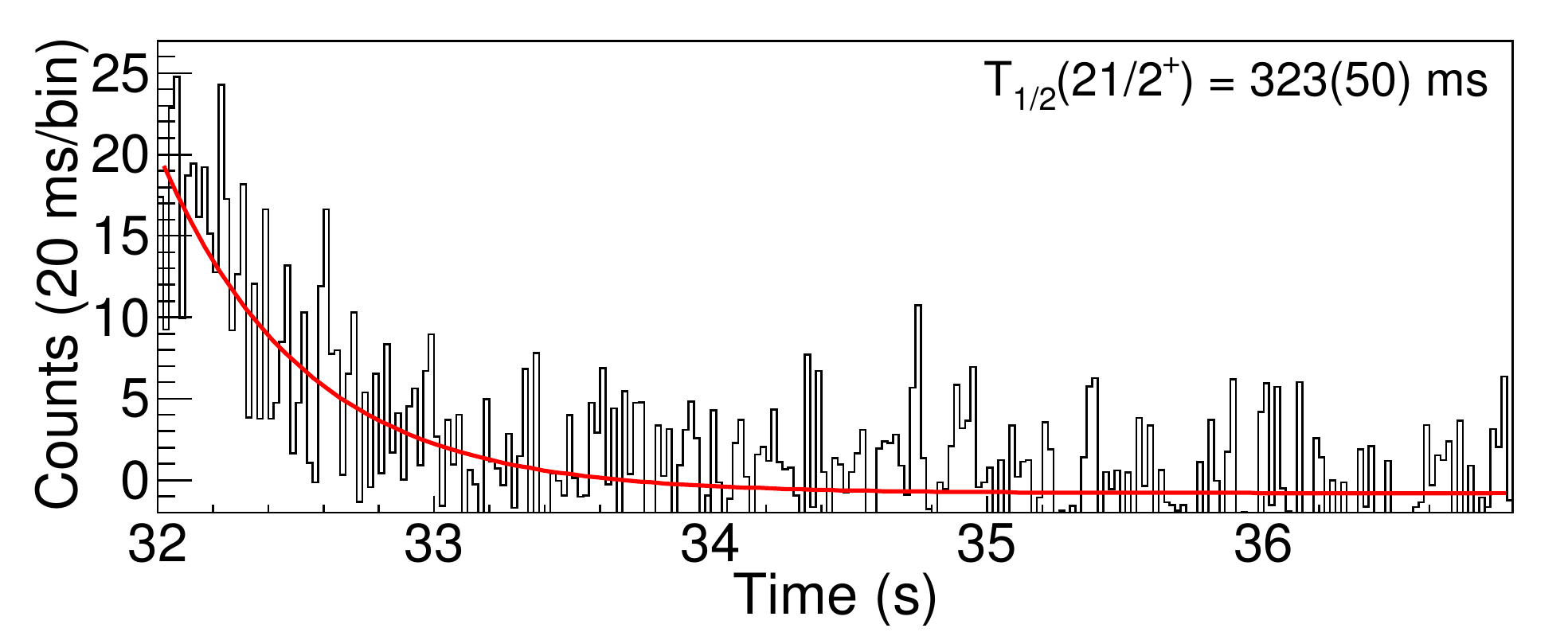}%
   \caption{(Color online) The fit of the activity of the 171, 173 and 284~keV $\gamma$ rays. These $\gamma$ rays from $^{131}$Sn are emitted following the $\beta$ decay of the $21/2^+$ state in $^{131}$In. The half-life from this fit was measured to be 323(50)~ms, in good agreement with the value of 320(60)~ms reported in Ref.~\cite{PhysRevC.70.034312}.
   \label{fig:21-2-halflife}}
\end{figure}

As shown in Fig.~\ref{fig:21-2-halflife}, the half-life of the $21/2^+$ $\beta$-decaying isomer of $^{131}$In was measured using a fit to the activity of the 171~keV, 173~keV and 284~keV $\gamma$ rays. The half-life of this state was measured to be 323(50)~ms, in good agreement with the value of 320(60)~ms reported in Ref.~\cite{PhysRevC.70.034312}.
\subsubsection{Decay of the $1/2^-$ State}
As there are no $1/2^-$ or $3/2^-$ states at low-excitation energy in $^{131}$Sn, the $\beta$-decaying $1/2^-$ isomeric state of $^{131}$In decays mainly by first-forbidden transitions to the $3/2^+$ ground state as well as the $1/2^+$ and $5/2^+$ low-lying excited states of $^{131}$Sn. There are also weak competing allowed $\beta$ decays to $1/2^-$ and $3/2^-$ states at excitation energies over 3.5~MeV, but the majority of the $\beta$ decay intensity is first-forbidden.

\begin{figure}[!t]
\centering
   \includegraphics[width=0.9\linewidth]{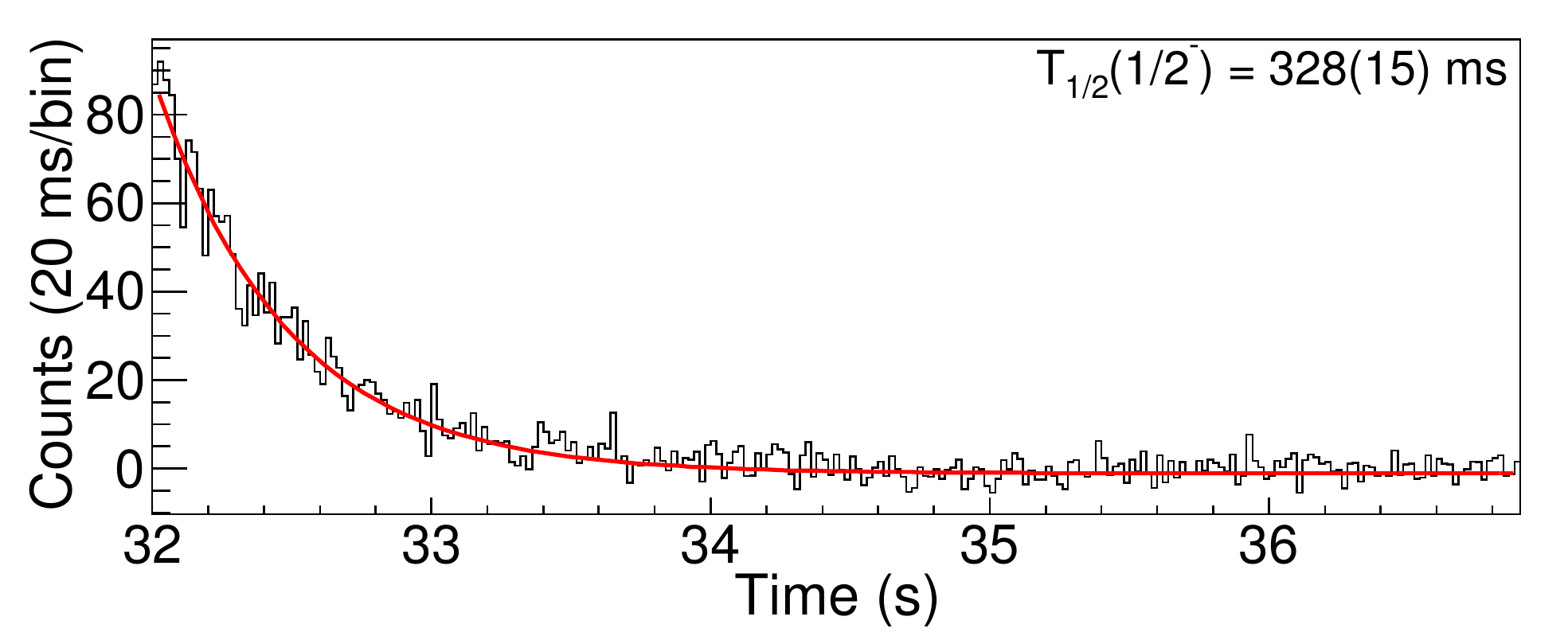}%
   \caption{(Color online) Fit of the half-life of the $1/2^-$ isomer in $^{131}$In. The activity curve shown was generated by placing a gate on the 332~keV $\gamma$ ray.
    The 332~keV $\gamma$ decay in $^{131}$Sb following the $\beta$ decay of the $^{131}$Sn daughter as well as the 330~keV $\gamma$ ray in $^{131}$Sn were accounted for. The half-life from this fit was measured to be 328(15)~ms, in good agreement with, but a factor of 3 more precise, than the value of 350(50)~ms reported in Ref~\cite{PhysRevC.70.034312}.
   \label{fig:1-2-halflife}}
\end{figure}

The half-life of the $1/2^-$ isomer was measured by fitting the activity of the 332~keV $\gamma$ ray from the decay of the $1/2^+$ state of $^{131}$Sn. The contribution of the $9/2^+$ $\beta$ decay that proceeds through the 332~keV $\gamma$-ray transition via a weak 1322~keV $\gamma$ ray branch from the 1654~keV state, as well as the 330~keV transition in the high-spin portion of the level scheme, were accounted for in the fit and had only a small effect on the measured half life. The activity of the 332~keV $\gamma$-ray in $^{131}$Sb following the $\beta$ decay of $^{131}$Sn was also accounted for in the fit. As shown in Fig.~\ref{fig:1-2-halflife}, using the 332~keV $\gamma$ ray, the half-life of the $1/2^-$ state was measured to be 328(15)~ms, in good agreement with, but a factor of 3 more precise than, the value of 350(50) reported in Ref.~\cite{PhysRevC.70.034312}.

\begin{figure}[t!]
    \centering
   \includegraphics[width=0.55\linewidth]{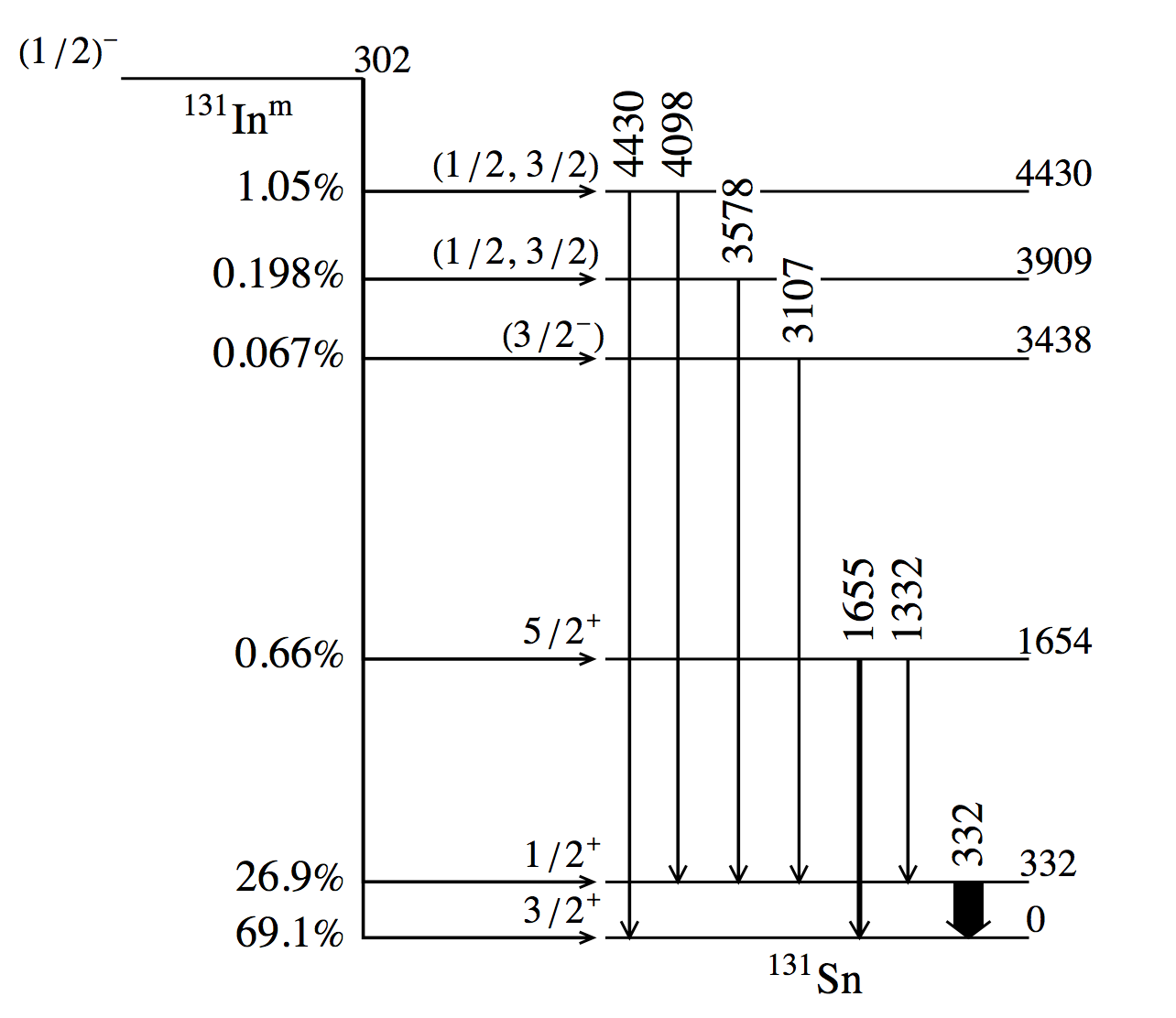}%
   \caption{Decay scheme of $^{131}$Sn observed following the $\beta$ decay of the $1/2^-$ state in $^{131}$In. The widths of the arrows represents the relative intensities of each of the $\gamma$-ray transitions.
   \label{fig:sn131-low-spin}}
\end{figure}

As shown in Fig.~\ref{fig:sn131-low-spin}, a much larger $\beta$ branch to the $1/2^+$ in $^{131}$In was deduced compared to previous works~\cite{PhysRevC.70.034312,fogelberg84}. In Ref.~\cite{PhysRevC.70.034312}, the $\beta$ feeding to the 332~keV excited state was reported to be 3.5(9)\%, compared to 26.9(6)\% measured in the current work. This increased feeding reduces the log$ft$ for this decay from 6.5 to 5.5. The predicted log$ft$ value of 5.74~\cite{zhi13} is in much better agreement with this new result. It is important to note that this transition is used in Ref.~\cite{zhi13} to determine the quenching factors for the first-forbidden operators in half-life calculations in the region. However, increasing the first-forbidden strength for the decay of $^{131}$In should result in an even shorter calculated half-life for $^{131}$In, further increasing the discrepancy between the calculated and experimental half-lives for $^{131}$In. 
\subsubsection{Decay of the $9/2^+$ State}
The $9/2^+$ ground state of $^{131}$In decays primarily by an allowed $\beta$ transition to the $7/2^+$ state of $^{131}$Sn at an excitation energy of 2434~keV. The 2434~keV state decays primarily by a strong $E2$ transition to the ground state. The only other potentially significant $\beta$ branch is the first-forbidden decay to the low-lying $11/2^-$ state in $^{131}$Sn.

\begin{figure}[!t]
\centering
   \includegraphics[width=0.9\linewidth]{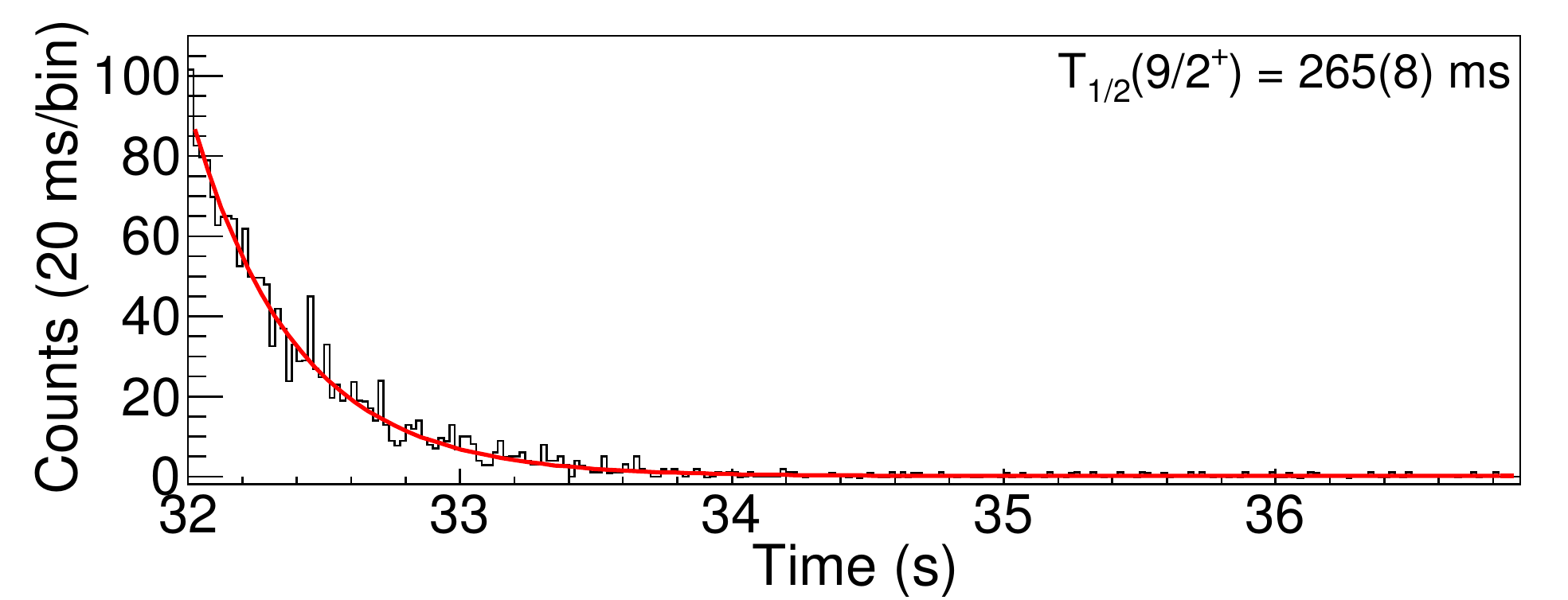}%
   \caption{(Color online) Fit of the half-life of the $9/2^+$ ground state of $^{131}$In from the activity of 2434~keV $\gamma$ ray. The half-life from this fit was measured to be 265(8)~ms, in good agreement with the most precise measurement of 261(3)~ms~\cite{lorusso15}.
   \label{fig:halflife-2434}}
\end{figure}

Figure~\ref{fig:halflife-2434} shows the fit of the activity of the 2434~keV $\gamma$-ray during the decay portion of the cycle. The half-life was measured to be 265(8)~ms, in good agreement with the most precise previous measurement of 261(3)~ms~\cite{lorusso15} as well as the measurement of 280(30)~ms~\cite{PhysRevC.70.034312}.

\section{Conclusions}
The new half-life measurements of $^{130}$Cd~\cite{lorusso15,PhysRevC.93.062801} are important as they imply a reduced quenching of the GT operator and resolved many of the discrepancies in the calculated half-lives for the neutron-rich $N=82$ nuclei important in the $r$-process. However, this result creates a new discrepancy for the half-life of $^{131}$In.

Strong first-forbidden $\beta$ feeding was observed from the $1/2^-$ state in $^{131}$In to the 332~keV $1/2^+$ state in $^{131}$Sn. This is in contrast to the 3.5(9)\% branching ratio reported in Ref.~\cite{PhysRevC.70.034312}. As the $\beta$ decay of the $1/2^-$ isomer is used in the scaling of first-forbidden $\beta$ decay strengths~\cite{zhi13}, future shell model calculations in the region should take this revision into account.

The half-lives of the $21/2^+$, $1/2^-$ and $9/2^+$ states in $^{131}$In measured using GRIFFIN are  323(50)~ms, 328(15)~ms and 265(8)~ms, respectively. Each of these half-lives is significantly longer than the half-life predicted by the shell-model calculations of Ref.~\cite{zhi13}. Further investigation into the discrepancy between the calculated and measured half-lives is therefore required.




%




\end{document}